\documentclass[aps,prapplied, twocolumn, superscriptaddress, showkeys]{revtex4-2}

\usepackage{graphics}
\usepackage{physics}
\usepackage{color}	
\usepackage{color,soul}	
\usepackage[dvipsnames]{xcolor}

\begin{document}


\title{Magnetization reversal driven by spin-transfer torque in perpendicular shape anisotropy magnetic tunnel junctions}


\author{N. Caçoilo}
\email[nuno.cacoilo@cea.fr]{}
\affiliation{Univ. Grenoble Alpes, CEA, CNRS, Grenoble-INP, SPINTEC, 38000 Grenoble, France}
\author{S. Lequeux}
\affiliation{Univ. Grenoble Alpes, CEA, CNRS, Grenoble-INP, SPINTEC, 38000 Grenoble, France}
\author{B. M. S. Teixeira}
\affiliation{Univ. Grenoble Alpes, CEA, CNRS, Grenoble-INP, SPINTEC, 38000 Grenoble, France}
\author{B. Dieny}
\affiliation{Univ. Grenoble Alpes, CEA, CNRS, Grenoble-INP, SPINTEC, 38000 Grenoble, France}
\author{R. C. Sousa}
\affiliation{Univ. Grenoble Alpes, CEA, CNRS, Grenoble-INP, SPINTEC, 38000 Grenoble, France}
\author{N. A. Sobolev}
\affiliation{I3N, Departamento de Física, Universidade de Aveiro, 3810-193, Aveiro, Portugal}
\author{O. Fruchart}
\affiliation{Univ. Grenoble Alpes, CEA, CNRS, Grenoble-INP, SPINTEC, 38000 Grenoble, France}
\author{I. L. Prejbeanu}
\affiliation{Univ. Grenoble Alpes, CEA, CNRS, Grenoble-INP, SPINTEC, 38000 Grenoble, France}
\author{L. D. Buda-Prejbeanu}
\affiliation{Univ. Grenoble Alpes, CEA, CNRS, Grenoble-INP, SPINTEC, 38000 Grenoble, France}

\date{\today}

\begin{abstract}
The concept of \textit{perpendicular shape anisotropy spin-transfer torque magnetic random-access memory} (PSA-STT-MRAM) consists in increasing the storage layer thickness to values comparable to the cell diameter, to induce a perpendicular shape anisotropy in the magnetic storage layer. Making use of that contribution, the downsize scalability of the STT-MRAM may be extended towards sub-20 nm technological nodes, thanks to a reinforcement of the thermal stability factor $\Delta$. Although the larger storage layer thickness improves $\Delta$, it is expected to negatively impact the writing current and switching time. Hence, optimization of the cell dimensions (diameter, thickness) is of utmost importance for attaining a sufficiently high $\Delta$ while keeping a moderate writing current. Micromagnetic simulations were carried out for different pillar thicknesses of fixed lateral size 20 nm. The switching time and the reversal mechanism were analysed as a function of the applied voltage and aspect-ratio (AR) of the storage layer. For AR $<$ 1, the magnetization reversal resembles a macrospin-like mechanism, while for AR $>$ 1 a non-coherent reversal is observed, characterized by the nucleation of a transverse domain wall at the ferromagnet/insulator interface which then propagates along the vertical axis of the pillar. It was further observed that the inverse of the switching time is linearly dependent on the applied voltage. This study was extended to sub-20 nm width with a value of $\Delta$ around 80. It was observed that the voltage necessary to reverse the magnetic layer increases as the lateral size is reduced, accompanied with a transition from macrospin-reversal to a buckling-like reversal at high aspect-ratios. 
\end{abstract}


\maketitle

\section{Introduction}
\label{sec:introduction}
The spin-transfer torque magnetic random-access memory (STT-MRAM) is one of the most promising emerging non-volatile memory technologies \cite{apalkov, khvalkovskiy2013, Song2018, gallagher2019}. It combines non-volatility with a quasi-infinite write endurance, high speed, low power consumption and scalability \cite {ikeda2010, carvello, Song2016}. These properties have stimulated the commercial production of STT-MRAM for a variety of standalone and embedded applications, in particular e-NOR FLASH replacement and last level Cache SRAM replacement \cite{Song2016, lee2018, dong2019, golongzka2018, slaughter2012}. While initial STT-MRAM devices used an in-plane (IP) magnetization, it has been shown that a perpendicular orientation of the magnetization leads to a better trade-off between the thermal stability factor $\Delta = \frac{E_\textrm{B}}{k_\textrm{B}T}$ (where $E_B$ is the energy barrier of the storage layer and $k_\textrm{B}T$ is the thermal activation energy with $k_B$ being the Boltzmann constant and T the operating temperature) which determines the memory retention time, and the switching current. These devices called perpendicular STT-MRAM (p-STT-MRAM) use the interfacial perpendicular magnetic anisotropy (iPMA) existing at the interface between the FeCoB layer and the MgO tunnel barrier\cite{apalkov, ikeda2010, dieny2017}. Nonetheless, downsize scalability of magnetic tunnel junctions (MTJ) below sub-20 nm diameters faces a fundamental challenge.  Indeed, as the device lateral size shrinks, there is a decrease in $\Delta$ due to the decrease in the iPMA total energy proportionally to the cell area. This can be understood considering that at these small dimensions, the reversal of the magnetic volume is almost coherent, and so $\Delta$ is proportional to the device area. This decrease significantly reduces the retention time of the memory \cite{yoshida2019, sato2017, thomas2014}. A proposal to counter this decrease is to double the iPMA by using two FeCo(B)/MgO interfaces \cite{sato2012}. Still, it remains challenging to keep $\Delta$ $>$ 80 at sub-20 nm diameters. A different approach consists in taking advantage of the shape anisotropy of the storage layer by increasing its thickness (\textit{L}) to values of the order or larger than the storage layer diameter. Thereby, the shape anisotropy no longer promotes an easy-plane magnetization but rather further stabilizes the magnetization in the perpendicular orientation. This concept of memory, named perpendicular shape anisotropy STT-MRAM (PSA-STT-MRAM), had been studied and experimentally developed for the first time simultaneously by SPINTEC \cite{perrissin2018, perrissin2019, nuno2019, steven2020, steven2021} and Tohoku University \cite{watanabe2018}. As the thickness of the storage layer increases, the magnetization is expected to be more stable due to the larger total anisotropy \cite{perrissin2018, perrissin2019, nuno2019, steven2020, watanabe2018} thus allowing to reduce the cell diameter beyond 20nm while maintaining a sufficiently large thermal stability. Another notable advantage of PSA-STT-MRAM is the much reduced sensitivity of their properties to the operating temperature \cite{steven2020}. This is due to the fact that the magnetic properties of their thick storage layer (in particular the thermal variation of its magnetization) are much closer to those of the bulk material than in conventional p-STT-MRAM, wherein the storage layer is only 1.5 nm to 2 nm thick. Thermal magnetic fluctuations can much more easily develop in such thin layers than in thicker ones.
However, the increased aspect-ratio (AR) of the storage layer in PSA-STT-MRAM may lead to a non-coherent reversal, higher switching voltages and longer switching times. For this purpose, the understanding of the magnetization reversal mechanism is necessary. In this work, micromagnetic simulations of the magnetization reversal mechanism of PSA-MTJs induced by STT were realised, enabling the identification of different features of the magnetization reversal. This study was realised on a pillar with fixed width of 20 nm and AR between 0.8 and 3. In a subsequent study, the width of the pillar was reduced to sub-20 nm dimensions, with an AR selected in order to maintain $\Delta$ around 80. In both studies, the reversal mechanism and the dependency between the switching time and the applied voltage was analysed. 

\section{Micromagnetic Model}
\label{sec:Model}
\begin{figure}[h]
\includegraphics{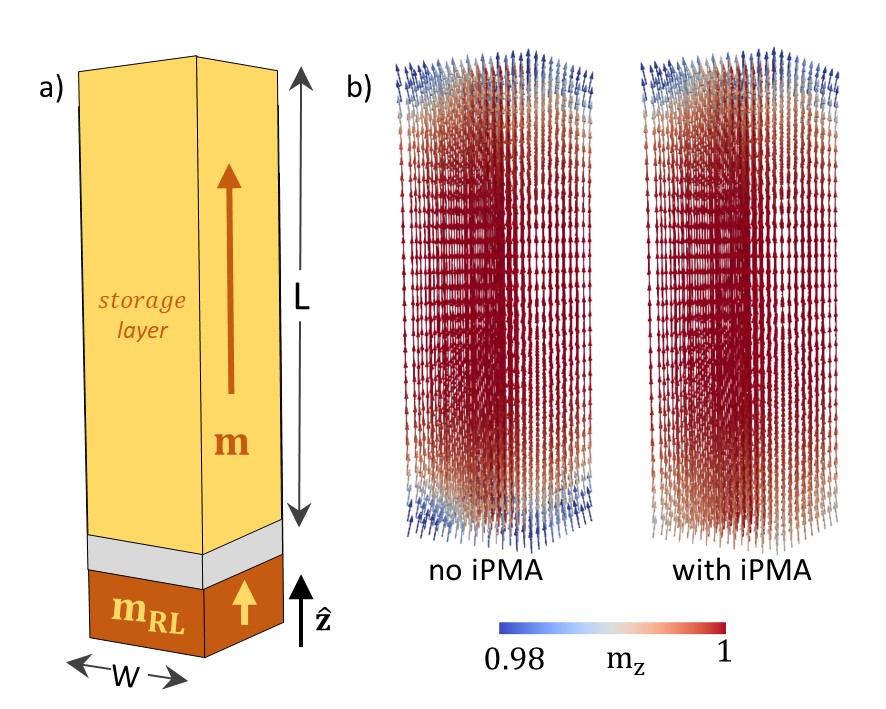}
\caption{a) 2D scheme of the studied FeCo(B) pillars with thickness L and base width W. The storage layer is shown with a yellowish colour, the tunnel barrier with a grey colour and the reference layer with a reddish colour. b) 3D equilibrium initial state of a 60 nm thick FeCo(B) layer without (left side) and with (right side) iPMA. The colour bar indicates the normalized magnitude of the magnetization along the defined z direction (along the pillar axis) in each cell.}
\label{fig:1}
\end{figure}

The magnetization dynamics is described by the Landau-Lifshitz-Gilbert-Slonczewski (LLGS) equation. We consider the material FeCo(B) (Fig.  \ref{fig:1}), with a spontaneous magnetization $M_s = 1$ MA/m, an exchange stiffness $A_{\textrm{ex}} = 15$ pJ/m and a damping value $\alpha$ of 0.01 \cite{DienyBook}. The simulations were performed using the finite-differences micro3D solver with a cubic cell size $\delta_\textrm{z}$ of 2 nm$^3$ \cite{micro}. The effective magnetic field $H_\textrm{{eff}}$ is calculated for each cell element. The iPMA is implemented numerically using an evanescent uniaxial contribution: 
\begin{equation}
K_u(z) = K_0\exp{-\frac{z}{\lambda_{\textrm{Ku}}}},
\end{equation}
where $\lambda_\textrm{Ku}$ defines the decay length of the iPMA throughout the thick layer. The value of the coefficient $K_0$ is adjusted so that the incremental sum of the areal value of $K_\textrm{u}$ results in $K_\textrm{s}$:
\begin{equation}
K_\textrm{s} = K_0 \sum_{i=0}^{N-1} \delta_\textrm{z} \exp \left( - \frac{i\delta_\text{z}}{\lambda_\textrm{Ku}}\right)
\label{eq:ko}
\end{equation}
with N the number of cell layers along the defined evanescent orientation.  

In Fig. \ref{fig:1} b) the equilibrium states of a magnetic pillar, with and without the effect of the interfacial anisotropy, are compared for a pillar with a squared base width of 20 nm and a thickness of 60 nm.  From Fig. \ref{fig:1} b) (left panel), a flower state is observed at both the top and bottom surfaces \cite{schabes1988}. When adding an iPMA at the bottom interface, a sturdier perpendicular orientation of the magnetization is enforced, Fig.  \ref{fig:1} b) (right panel). 

Starting from the equilibrium state with iPMA, a spin-polarized current is injected. In a thin MTJ, this effect is included in the LLGS equation as a damping-like torque term ($\Gamma_\textrm{STT}^\textrm{IP}$) and a field-like torque (the latter neglected in the following equations) \cite{apalkov}:
\begin{equation}
\partial_t\mathbf{m} = -|\gamma|\mu_0\left(\mathbf{m}\times\mathbf{H_\textrm{eff}} \right) + \alpha\left(\mathbf{m}\times\partial_t\mathbf{m} \right) +  \Gamma_\textrm{STT}^\textrm{IP},
\end{equation}
where $\gamma$ is the gyromagnetic ratio, $\mu_0$ the vacuum permeability and: 
\begin{equation}
\label{eq:STT}
\Gamma_\textrm{STT}^\textrm{IP} = - |\gamma| a_\parallel V \mathbf{m}\times(\mathbf{m}\times \mathbf{m_\textrm{RL}})
\end{equation}
where $a_\parallel$ is the pre-factor of the damping-like torque, \textit{V} the applied voltage, $\mathbf{m}$ the normalized magnetization vector of the storage layer and $\mathbf{m_\textrm{RL}}$ the normalized magnetization vector of the reference layer. 

As the injection of current occurs at the bottom interface, the spin polarization is expected physically to decay exponentially through the interaction with the magnetization \cite{chshiev, grollier2011}. We can model this effect assuming that the value of $a_\parallel$ decreases spatially as:
\begin{equation}
a_\parallel = a_0\exp{-\frac{z}{\lambda_\textrm{STT}}},
\end{equation}
where $\lambda_\textrm{STT}$ defines the length scale of the STT decay. The value of the coefficient $a_0$ can be obtained knowing that the averaged sum of each plane of cells along the material will result in the macrospin value of $\langle a_\parallel \rangle$:
\begin{equation}
\langle a_\parallel \rangle = \frac{\hbar}{2|e|}\frac{\eta_\textrm{STT}}{RA}\frac{1}{M_\textrm{s}L}
\label{eq:apara}
\end{equation}
with $\hbar$ the reduced Planck constant, $e$ the elemental charge, $\eta_\textrm{STT}$ the STT efficiency, $RA$ the resistance-area product of the tunnel barrier and $L$ the total thickness of the storage layer. From this value, it is possible to calculate the coefficient $a_0$ in the following way: 
\begin{equation}
\langle a_\parallel \rangle = \frac{a_0}{N}  \sum_{i=0}^{N-1} \exp \left(-\frac{i\delta_\textrm{z}}{\lambda_\textrm{STT}}\right).
\label{eq:ao}
\end{equation}
In the present work, both decay lengths are assumed to be of 1 nm. In addition, considering the very small area of the pillar, operable PSA-STT-MRAM pillars require a \textit{RA} product of the order of 1 $\Omega \cdot \mu m^2$ to avoid excessive write voltages which may yield dielectric breakdown. Since the STT efficiency is a function of the tunnel magnetoresistance (TMR) \cite{sun_spin-torque_2013, slonczewski_theory_2007, sun_magnetoresistance_2008}, and assuming that in our device the TMR is higher than 100 $\%$, a value of $\eta_\textrm{STT} = 0.5$ is used. 

\section{Magnetization reversal driven by STT}

\begin{figure}[h]
\includegraphics{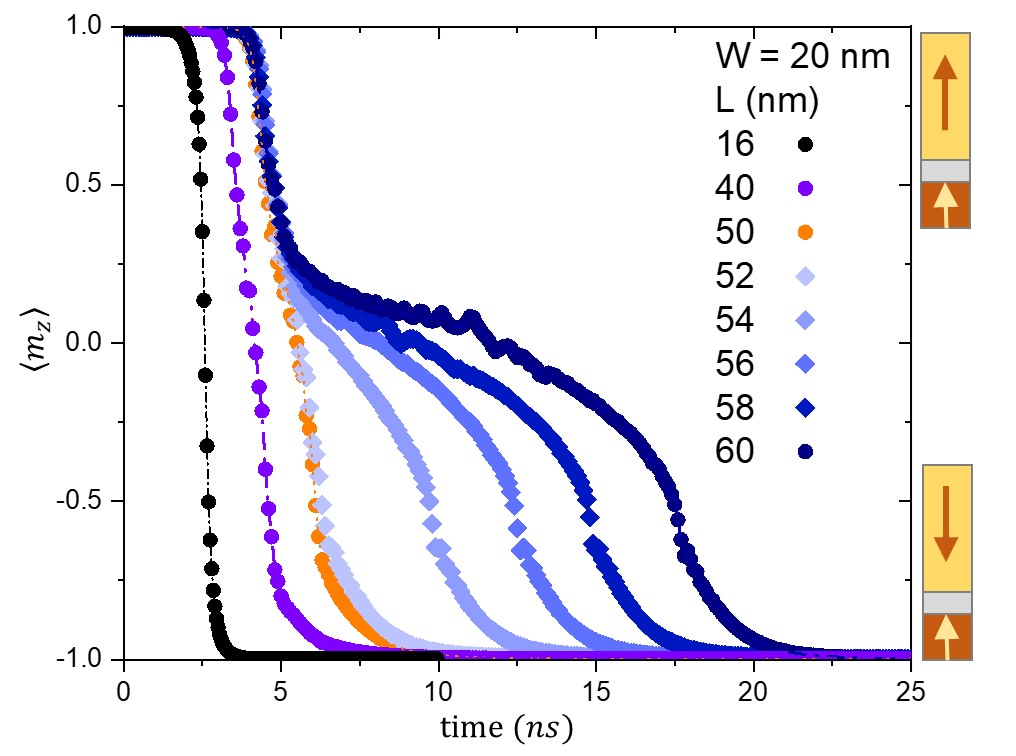}
\caption{Time traces of the mean reduced magnetization $\langle m_z \rangle$ for a storage layer with an AR between 0.8 and 3, for a fixed width of W = 20 nm, at an applied voltage of -3 V}
\label{fig:2}
\end{figure}

We consider hereafter a set of magnetic pillars with constant width of 20 nm and different pillar thicknesses (L = 16, 18, 20, 30, 40, 50 and 60 nm). Figure \ref{fig:2} shows the average magnetization along the vertical axis (growth direction) of the storage layer as a function of time for the different thicknesses for an applied voltage of -3 V. This voltage range is actually experimentally inaccessible since it would cause the dielectric breakdown of the tunnel barrier. However, one could reduce this voltage range by lowering the junction \textit{RA} product to values much below 1 $\Omega \cdot \mu m^2$ (as is the case in recording heads) at the expense of a lower TMR. It is observed that, for the same voltage, an increase in thickness results in an increase in the time it takes to start reversing the magnetic layer. While for a thickness of 16 nm, the switching mechanism exhibits a rather sharp variation of the magnetization during reversal, a slower relaxation is obtained for the thicker layers (see \textit{e.g.} 40 and 50 nm in figure \ref{fig:2}). Further increasing the thickness of the storage layer leads to a change in the reversal profile. Around 52 nm, the magnetization reversal starts exhibiting a shoulder, an effect that worsens as the layer thickness is increased up to 60 nm.

\begin{figure}[h]
\includegraphics{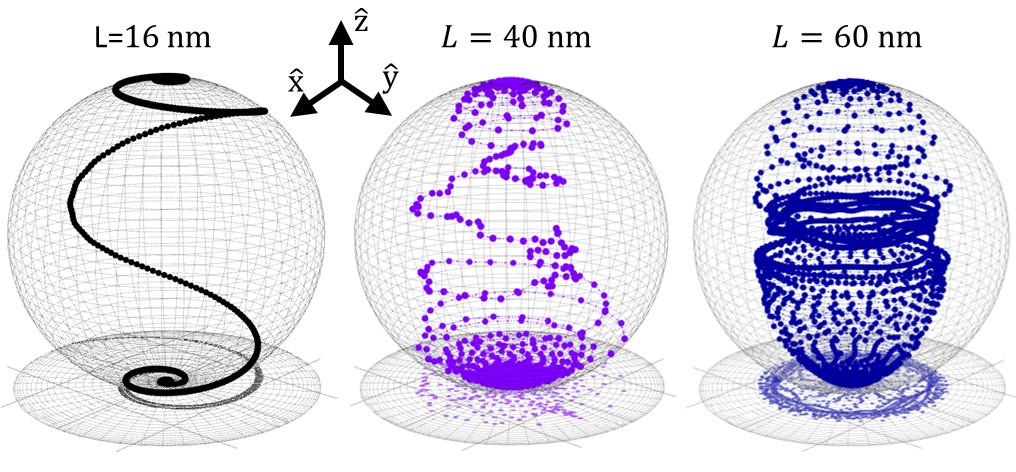}
\caption{Switching trajectories for layers of different thicknesses. The z-axis represents $\langle m_z \rangle$ and the basal plane (x and y axis) represents  $\langle m_{x,y} \rangle$. The simulated data is contained in a macrospin sphere (radius 1). Results obtained for an applied voltage of -3 V for a pillar thickness of 16, 40 and 60 nm width a fixed width of W = 20 nm.}
\label{fig:3}
\end{figure}

To understand the underlying mechanism of reversal in these structures, we can resort to the 3D trajectories described by the mean magnetization vector inside the unitary sphere, shown in figure \ref{fig:3}. 

\begin{figure}[h]
\includegraphics{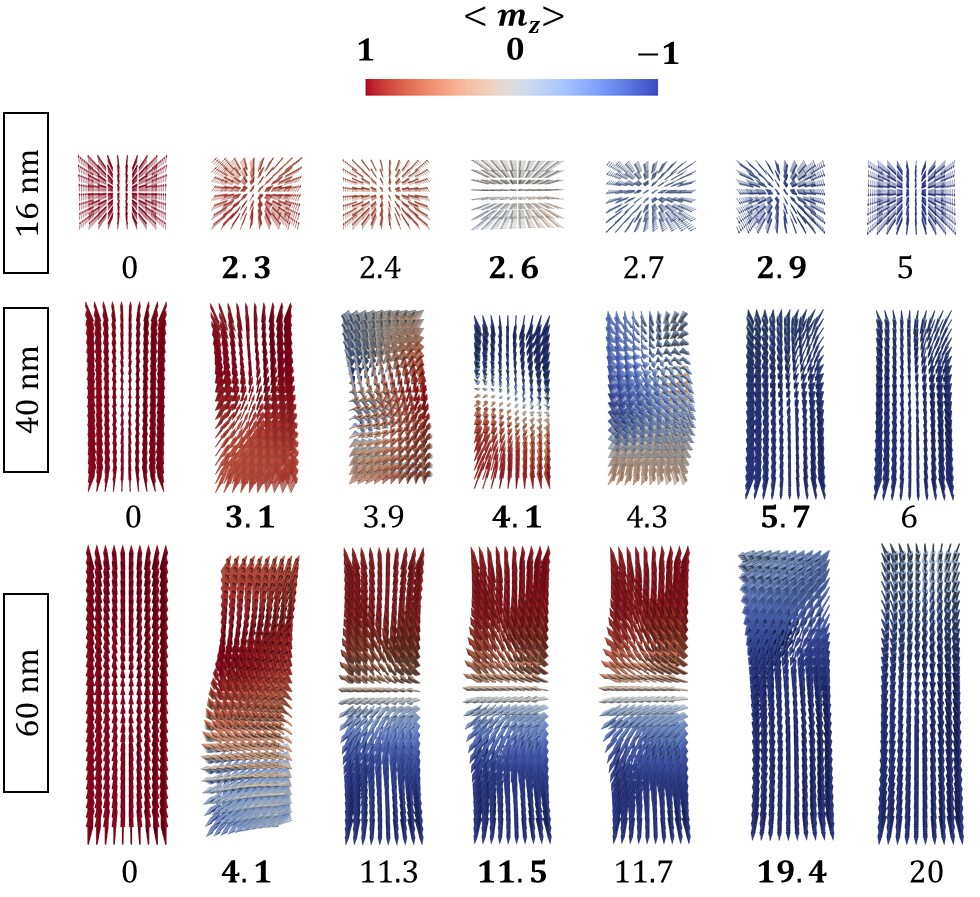}
\caption{Snapshots at different time steps (at bold is shown the time were we reverse 10$\%$, 50$\%$ and 90$\%$ of the magnetic layer)  for a 16 nm, 40 nm and 60 nm (at an applied voltage of - 3 V) thick magnetic layer with a width of 20 nm. The color is representative of the magnitude of $\langle m_z \rangle$ and quantified in the color bar.}
\label{fig:4}
\end{figure}

Indeed, from these trajectories, we can infer that the switching mechanism is thickness-dependent: for a 16 nm thickness, a typical macrospin switching trajectory is observed since the normalized magnetic moment keeps its maximum amplitude throughout the magnetic reversal. When increasing the thickness to 40 nm, the switching trajectory no longer lies on the unitary sphere. This effect is more pronounced for a thickness of 60 nm, combined with a slowing down of the dynamics. This slowing down was already observed in figure \ref{fig:2}, related with the shoulder near $\langle m_z \rangle~\approx~0$. We can further observe this behaviour from the 3D snapshots of the magnetization, shown in figure \ref{fig:4} at different time steps, for the 3 different thicknesses, 16 nm, 40 nm and 60 nm. Starting with the lower thickness of 16 nm: the macrospin regime is identified, the magnetization rotates coherently. Even though the model assumes the STT to decay exponentially from the MgO interface, at these dimensions, the exchange interactions are strong enough to insure coherent reversal. For the thickness of 40 nm, the magnetization reversal follows a buckling-like mechanism since the whole magnetic layer reacts simultaneously during the reversal but in a non-uniform way\cite{aharoni1988, aharoni1958, hertel2004}. For the larger thickness of 60 nm, the nucleation of a domain wall starts at the bottom interface (where the STT is being transfered). Two magnetic domains, in a tail-to-tail domain wall configuration are observed in the frame at 11.5 ns. The domain wall then propagates along the vertical direction of the magnetic layer, while it rotates azimuthally in the transverse plane (as seen for example in figure \ref{fig:3}). This mechanism of reversal is thus identified as a transverse domain wall propagation \cite{thiaville2005}.

\begin{figure}[h]
\includegraphics{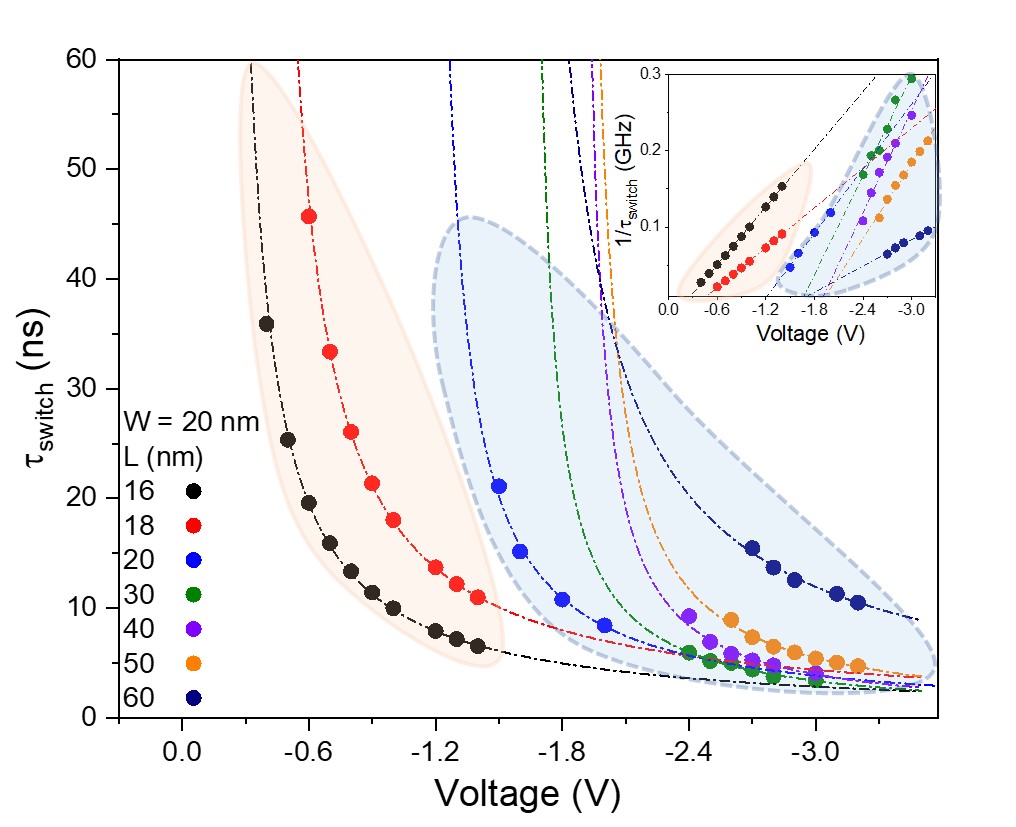}
\caption{Dependency of the switching time on the applied voltage for the different AR values studied. The coherent reversal is highlighted with a pinkish watermark and the non-coherent with a blue watermark. The inset shows the  dependency of the inverse of the switching time on the applied voltage.}
\label{fig:5}
\end{figure}

From the dependencies observed in figure \ref {fig:2}, it is possible to extract an important feature of the STT-driven reversal, the dependency of the switching time on the applied voltage. This switching time is defined as the time it takes to reverse half of the magnetic volume - $\tau_{\textrm{switch}}$ - which, in a macrospin regime, is given when the magnetization lies perpendicular to its initial orientation, \textit{i.e.} $\langle m_z \rangle = 0$. This is shown in figure \ref{fig:5}, with an inset showing the dependency of the inverse of $\tau_{\textrm{switch}}$ versus applied voltage. This relationship is found to be linear, which is a consequence of the conservation of angular momentum during the reversal process \cite{garello2014, liu2014, worledge2011}. Moreover, the slope of this linear behaviour is related to the magnitude of the perpendicular anisotropy field ($H_{\perp}$) and to $\langle a_\parallel \rangle$ (using equation \ref{eq:apara}) \cite{DienyBook}:
\begin{equation}
\frac{1}{\tau_\textrm{switch}} = V \left(\frac{\langle a_\parallel \rangle^2}{H_{\perp}}\right)\mathcal{A} - \langle a_\parallel \rangle\mathcal{B}.
\end{equation}
with $\mathcal{A}$ and $\mathcal{B}$ thickness independent constants.  

Indeed, fixing the diameter and increasing the thickness leads to an increase in $H_\perp$, as it is dominated by the contribution of the PSA (the value of the iPMA is fixed for all the different thicknesses). In addition, with increasing thickness, $\langle a_\parallel \rangle$ is reduced (see equation \ref{eq:apara}). As a result, the slope becomes steeper as long as the reversal is coherent (AR of 0.8 and 0.9). Further increasing the thickness of the storage layer leads to a different evolution of the slope, which reflects the non-coherent reversal behaviour. This was already expected from figure \ref{fig:2}, as there is a clear deviation from the macrospin regime. 

Moreover, it is observed that an increase in thickness is accompanied by an increase in the minimum voltage to reverse the magnetic layer, which, as will be shown below, is due to an increase in the thermal stability factor. This means that at this width, an AR much higher than 1 would not be practical for a working device, because of the high switching voltage and slow reversal mechanism.

\begin{figure}[h]
\includegraphics{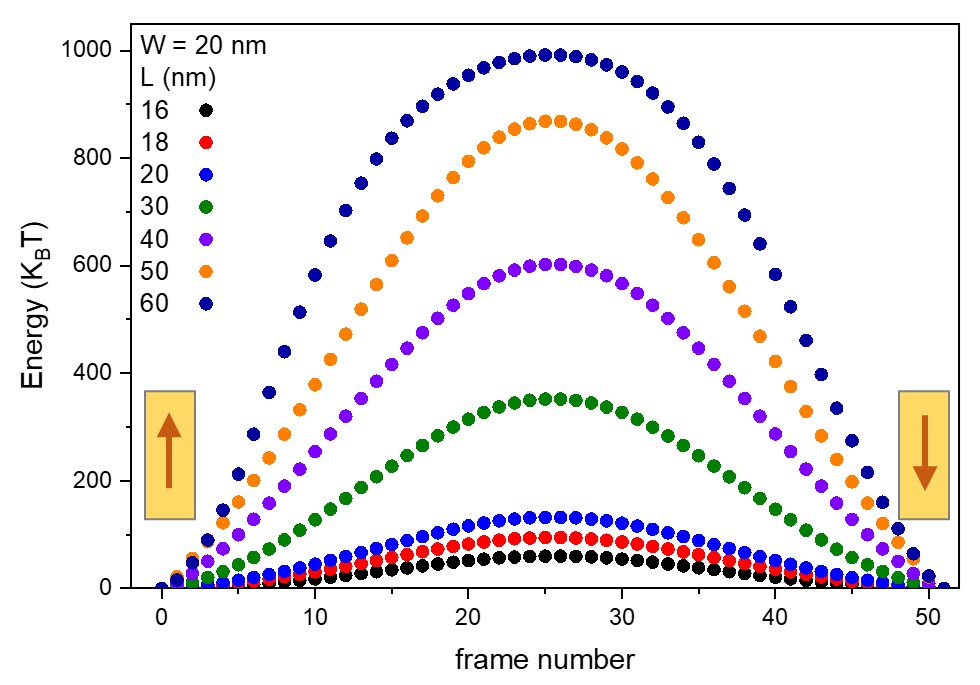}
\caption{Minimum energy path of the energy barrier for different thicknesses of the storage layer with a fixed diameter of 20 nm. }
\label{fig:6}
\end{figure}

The thermal stability can be calculated using the string method \cite{Forster2003, Weinan2003, Chaves2009, Chaves2010, Carilli2015}, which computes the minimum energy path (MEP) that the magnetization follows to reverse between its two stable states (magnetization up or down), shown in Fig. \ref{fig:6}. As expected, increasing the thickness of the storage layer significantly increases the energy barrier, leading to an increase in $\Delta$, which rapidly gets prohibitively large for AR higher than 1. Thus, it is important to take into consideration the AR of the storage layer when designing the device, as $\Delta$ is very sensitive to the layer thickness.

\section{STT-driven reversal in sub-20 nm width MTJ with $\Delta\approx 80$}

From the perspective of device applications, one should stay within a reasonable value of $\Delta$ (typically in the range 60-100 depending on the memory specified retention and acceptable bit error rate). This is the case for an AR $\lesssim 1$ at these selected material parameters. Staying within this AR also leads to a coherent reversal,  avoiding the longer switching time and higher voltages needed for the high AR pillars. As such, we studied different AR combinations giving a $\Delta$ of around 80, with dimensions (both width and thickness) in the sub-20 nm design space.  For these simulations, as we are using smaller thickness than before, the cell size was reduced from 2 nm to 1 nm. The selected dimensions of the storage layer comprise widths of 18, 16, 14, 12 and 10 nm, with an adjusted AR of 0.9, 1, 1.1, 1.3 and 1.8, respectively. 

\begin{figure}[h]
\includegraphics{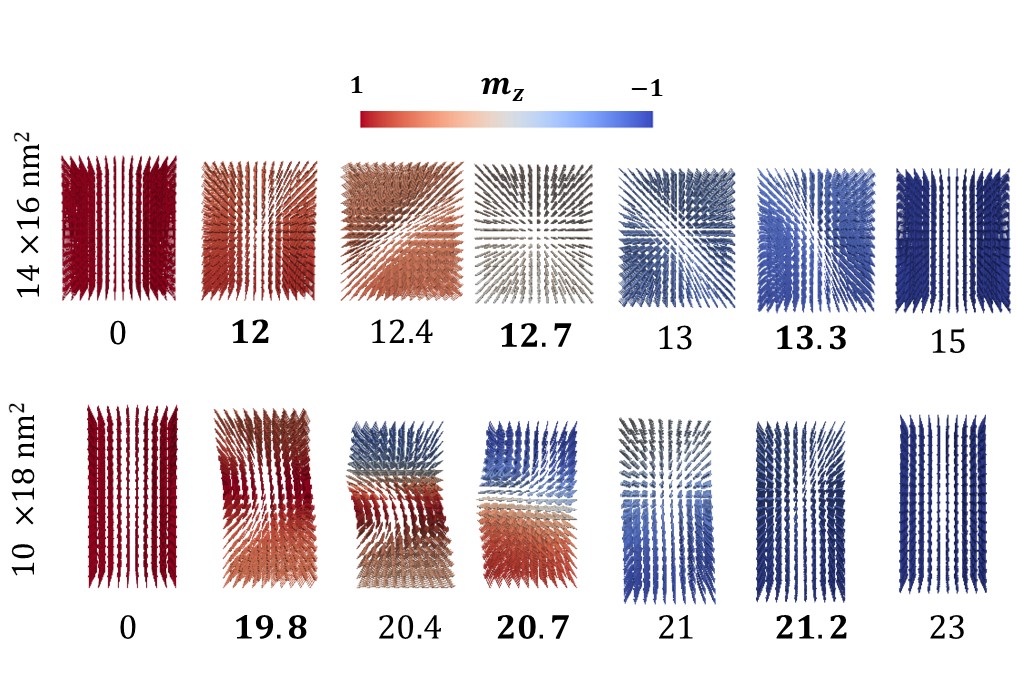}
\caption{Snapshots at different time steps (in bold are shown $\tau_{10\%}$, $\tau_{50\%}$ and $\tau_{90\%}$ in ns)  for a pillar of 14 nm width and 16 nm thickness (at an applied voltage of - 1 V) and a pillar of 10 nm width and 18 nm thickness (at an applied voltage of - 1.4 V). The color is representative of the magnitude of $\langle m_z \rangle$ and quantified in the color bar.}
\label{fig:7}
\end{figure}

Even though $\Delta$ is the same, it is possible to observe two different reversal mechanisms in the 3D snapshots of the magnetization, shown in figure \ref{fig:7} at different time steps. It is observed that if we keep an AR near 1, the magnetization reversal is coherent, while for a larger AR, non-coherent reversal occurs. 

\begin{figure}[h]
\includegraphics{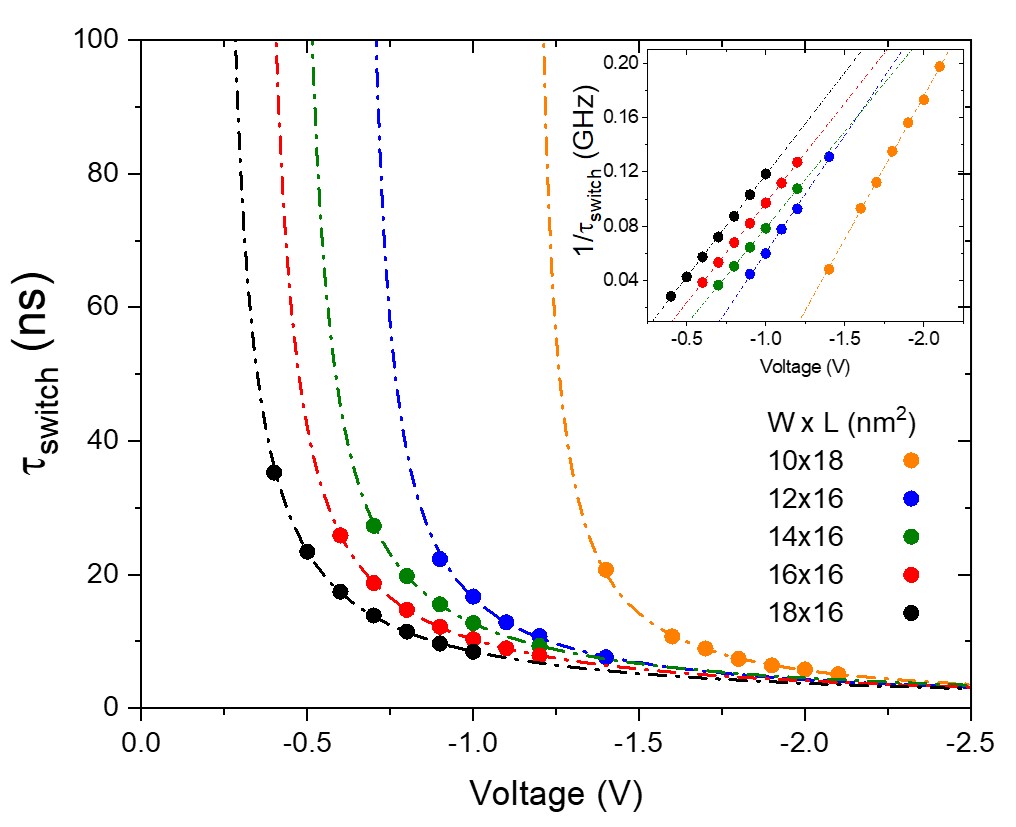}
\caption{Dependency of the the switching time with the applied voltage for the different studied AR. The inset shows the  dependency of the inverse of the switching time on the applied voltage.}
\label{fig:8}
\end{figure}

The dependency of the switching time on the applied voltage is shown in figure \ref{fig:8}, with the inset showing the conservation of the linear behaviour between $\tau_{\textrm{switch}}^{-1}$ and the applied voltage. By increasing the AR, the voltage required to reverse the magnetization in the storage layer increases, even though the thermal stability factor is fixed. In addition, there is no clear variation in the slope for the AR's of 0.9, 1 and 1.1, which can be attributed to the similar value of $\langle a_\parallel \rangle$ and the weak variation of the PSA (in comparison with the significant variation in AR seen in figure 5). When increasing the AR from 1.1 to 1.3 and 1.8, we can see a difference in slope, related to a non-uniform reversal. Indeed, at these dimensions, the reversal is no longer macrospin and the nature of the reversal is strongly linked to the dipolar coupling created by these low width pillars, leading to a buckling-like reversal, as observed in figure \ref{fig:7}. 

This allows us to conclude that the type of magnetization reversal is dependent on the AR and not on the $\Delta$ itself. In a first approach, it seems difficult to avoid the non-coherent reversal that accompanies the high AR required to maintain the perpendicular anisotropy when going to sub-20 nm. A possible solution to this challenge was experimentally presented by B. Jinnai \textit{et. al.} \cite{Jinnai1, Jinnai2}, in which a thin non-magnetic spacer layer is placed in the middle of the storage layer. With this design, it is no longer energetically favourable for a non-coherent reversal to occur, leading to the possibility of having a coherent-reversal in sub-20 nm PSA-STT-MRAM.  

\section{Conclusion}

Using micromagnetic calculations, we have investigated the magnetization reversal in the PSA-MTJ at sub-20 nm width. For a fixed width of 20 nm, it was observed that for the pillars with smaller AR, the reversal exhibits a macrospin-like behaviour. When increasing the thickness, a non-coherent reversal takes place, which evolves from a buckling-like reversal to a transverse-domain wall propagation for higher AR. The latter is associated with a slowing down of the dynamics when the domain wall is located around the middle of the storage layer. It was further observed that the inverse of the switching time follows a linear relationship versus applied voltage. 

Considering practical device applications, we maintained a thermal stability factor of around 80 and varied the width beyond sub-20 nm dimensions. As in the first study, the magnetization reversal evolved from macrospin to a non-coherent reversal with increasing AR. This allowed us to conclude that the type of magnetization reversal mainly depends on the AR and not on the $\Delta$ itself. Moreover, the linear dependency $\tau_\textrm{switch}^{-1} \propto V$ is maintained over the whole range of cell diameters investigated. 

This study provides some guidelines for the design of PSA-STT-MRAM, that can be summarized as follow: (1) the resistance$\times$area product , as used previously, must be lowered in comparison to that in conventional STT-MRAM to maintain the write voltage significantly below the barrier breakdown voltage. (2) The $\Delta$ should be maintained around a window of 60-100 depending on the memory retention and acceptable bit error rate specifications, which corresponds to an AR $\approx$ 1. Using two MgO interfaces instead of one is a possibility to further reduce the storage layer thickness. This allows to avoid excessive voltages and the slowing down introduced by the domain wall nucleation and propagation. (3) In macrospin reversal, the switching time can be reduced to values of the order of 20 ns - which makes this type of memory not ultrafast but still compatible with DRAM type of applications (for instance battery-backed DRAM), provided high memory densities can be obtained. In addition, besides the e-NOR FLASH replacement mentioned in the introduction, by lowering the RA product to sub- 1 $\Omega \cdot \mu m^2$ values, the switching time could be further reduced opening prospects of applications as SRAM for low power wearable systems, all sorts of micro-controllers and internet of things devices. 

\begin{acknowledgments}
This work was supported by Samsung Electronics Co., LTD. (IO190709-06540-02).
\end{acknowledgments}

\bibliography{references}

\end{document}